Title: Exciton matter sustained by colossal dispersive interactions due to enhanced polarizability: Possible clue to ball lightning
Authors: Mladen Georgiev (1), Jai Singh (2) ((1) Institute of Solid State Physics, Bulgarian Academy of Sciences, Sofia, Bulgaria, (2) School of Engineering and Logistics, Charles Darwin University, Darwin, Australia)
Comments: 9 pages, 2 figures, pdf format

Recently Gilman has pointed out that the material state of a ball lightning is both highly cohesive and flexible. He makes a specific proposal for a cohesive state arising from (colossal) Van-der-Waals attraction between highly polarizable Rydberg atoms produced under a linear lightning. We accept his general suggestions but propose that the colossal Van-der-Waals coupling may also arise from the enhanced polarisability of surrogate molecular clusters due to the polaron gap narrowing effect. We consider a few illuminating cases and present calculations for the ammonia molecule. Although being unable to identify the exact nature of the surrogate molecules at least for the time being, we suggest a general scenario of photoexcited vibronic excitons forming a supersaturated surrogate gas phase in which a ball arises as a result of condensation. The orange color of the luminous ball is due to radiative exciton deexcitation and suggests that there may be a unique surrogate material for ball lightning.

1. Introduction

The scientific interest in the excited state of matter has increased dramatically lately with relevance to the enigmatic ball lightning: A recently proposed model regards the ball as a highly excited state of an unspecified substance composed of Rydberg atoms excited by the vast discharge of a linear lightning.[1] These atoms being highly polarizable, they give rise to colossal dispersive interactions which keep the ball in a strongly cohesive though highly flexible state within a few seconds after the excitation. Yet the model could not explain the orange light emitted by the luminous ball as it deexcites. This light apparently originates from an intrinsic luminescence hardly attrubutable to Rydberg atoms.[2] One of the most popular theories nowadays seems to be the one attributing ball formation to the oxidation of silicon nanoparticles in the atmosphere as a linear lightning hits the ground.[3] The evolution of silicon nanoparticles is stimulated as carbon in the soil reduces the silicon oxides to metallic silicon at the high lightning-strike temperatures. As a matter of fact, traces of silicon have really been found at deexcitation sites as luminous plasma disks of apparently similar nature have been observed to form triggered by the detonation of explosives evolving silicon. Nevertheless, the situation seems more complicated since a ball has been observed to form on aircraft.[2] All this evidence seems to justify other attempts to add novel background to the unsolved ball-lightning problem.

Herein we would like to show that in addition to Rydberg atoms there is at least one more kind of highly polarizable species produced through photoexcitation. Some time ago we studied theoretically the colossal dispersive couplings arising from an enhanced polarizability of excitonic small-polaron species. The increased polarizability is obliged to the interplay of two factors: (i) an electric dipole due to inversion symmetry breaking and (ii) a polaron band narrowing (Holstein), both of them occurring in a lower-symmetry vibronic-polaron manifold.[4] The resulting vibronic polarizability might surpass by one to two orders of magnitude the normal electrostatic polarizability of a two-level system.[5] Consequently, Van-der-Waals dispersive interactions of unusual strength might appear in a two-level system coupled to an asymmetric vibrational mode.

It will be seen that while Gilman's quest is for systems with an enhanced dipole moment (Rydberg atoms) in order to elevate the *numerator* of the polarizability equation, we look for alternative systems (vibronic excitons) to reduce the *denominator*. The common result of both approaches is enhancing the polarizability of excited atomic clusters.

## 2. Binding energy of vibronic exciton matter

A vibronic exciton forms as the ultimate product of the relaxation of the band exciton (BE) to a self-trapped state (STE1) followed by a subsequent relaxation to another lower-lying self-trapped state (STE2) before producing a defect pair (DP). Normally the BE-STE1 step of the transition process is presumed coupled to a symmetric vibration. The vibronic effects appear as the STE1-STE2 step couples to an asymmetric vibration. It should be stressed that in the simplest case two nearly-degenerate different-parity electronic states associated with STE2 are needed to mix vibronically by an asymmetric odd-parity vibrational mode. The overall process is exemplified in Figure 1 showing the sea of band excitons as well as the adiabatic potentials of subsequent symmetric and asymmetric self-trapped states. Similar qualitative diagrams have been considered elsewhere to illustrate the transfiguration of excitons to defect pairs in alkali halides and silicon oxide.[6,7] What is essential is that the asymmetric-mode coupling lends an electric dipole moment to the STE2 state, due to the breaking of inversion symmetry. The vibronic dipole might vanish on the average if the asymmetric species performed flip flops or rotations around the higher-symmetry site which would in effect smear the dipole moment. With dipoles smeared or not, the breaking of inversion symmetry will give rise to dispersive interactions (*a la* London) within the STE2 ensemble. One is the vibronic extension of the Van-der-Waals coupling of atoms.

It should be stressed unequivocally at this point that the on-going conclusions are by no means dependent on whether the medium is crystalline or amorphous, a dense liquid, or is even in a soft aggregate state. What is needed is to have a manifold of similar molecular clusters, each cluster breaking the inversion symmetry in static state even though restoring it on the average in the dynamic rotatory state. An example familiar to chemists is the ammonia molecule with left-handed and right-handed configurations.[8]

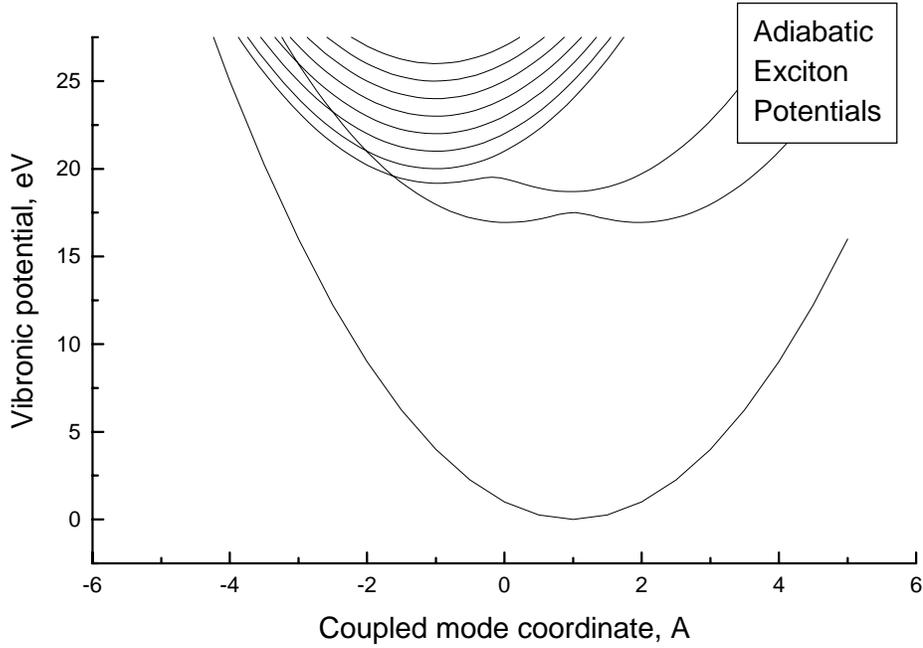

Figure 1.
Adiabatic vibronic potentials pertaining to polaronic excitons (schematic). Notations: GS – ground state, BS – band states, STE1 – symmetric self-trapped state, STE2 – asymmetric self-trapped state. The radiative transitions from BS, STE1 and STE2 follow vertical paths from the respective minima down to GS

Concomitantly we consider a molecular cluster with two nearly-degenerate opposite-parity excitonic eigenstates $\psi_1$ and $\psi_2$ of eigenenergies $\varepsilon_1$ and $\varepsilon_2$, respectively. If now $\mathbf{p}_{12} = <\psi_2|\,e\mathbf{r}\,|\psi_1>$ is the electric dipole mixing these states, the electrostatic polarizability of the two-level system will be

$$\alpha_{exc} = (1/3)\, p_{12}^2 / |\varepsilon_2 - \varepsilon_1| \qquad (1)$$

The two-level electrostatic polarizability (1) brings about a VdW exciton pairing energy

$$U_{VdWexc} = \tfrac{1}{2} E_{gap} (\alpha_{ex} / \kappa R_{ij}^3)^2 \qquad (2)$$

$\kappa$ is the dielectric constant of the medium, $R_{ij}$ is the pair separation, $E_{gap} = |\varepsilon_2 - \varepsilon_1|$ is the interlevel energy gap.

We subsequently assume that the two electronic states are mixed *vibronically* by an odd-parity intermolecular vibration Q through the *pseudo-Jahn-Teller effect*. As a result of the vibronic mixing the original interlevel spacing $E_{gap}$ reduces to

$$\Delta E = E_{gap} \exp(-2E_{JT}/\hbar\omega) \qquad (3)$$

where $E_{JT}$ is the Jahn-Teller energy, $\omega \equiv \omega_{ren} = \omega_{bare}\sqrt{[1 - (E_{gap}/4E_{JT})^2]}$ is the renormalized frequency of the coupled vibration, $\omega_{bare}$ being the bare vibrational frequency. The electrostatic polarizability of the squeezed two-level system turns into

$$\alpha_{vib}^0 = (1/3)\, p_{vib}^2 / |\Delta E|, \qquad (4)$$

termed *vibronic polarizability*, with

$$p_{vib} = p_{12}\sqrt{[1 - (E_{gap}/4E_{JT})^2]} \qquad (5)$$

standing for the *vibronic dipole*. The vibronic mixing effects give rise to a *vibronic-exciton VdW pairing energy* in lieu of equation (2):[4,5,9]

$$U_{VdWvib} = \tfrac{1}{2} \Delta E\, (\alpha_{vib}^0 / \kappa R_{ij}^3)^2 \qquad (6)$$

(See References [4] and [5] for details.) The vibronic polarizability $\alpha_{vib}$ is temperature dependent ($\alpha_{vib}^0$ is its low-temperature value).. For a molecular system:[8]

$$\alpha_{vib}(T) = \alpha_{vib}^0 \tanh(|\Delta E|/k_B T) \qquad (7)$$

From (6) we obtain the electrostatic *Van-der-Waals binding energy* $U_{vib}$ of a STE2 vibronic-exciton system:

$$U_{vib} = \tfrac{1}{2} \Delta E\, (\alpha_{vib}^0/\kappa)^2 \sum R_{ij}^{-6} \qquad (8)$$

The vibronic energy (8) is to be compared with the *VdW binding energy* $U_{ex}$ of a system of excitons in STE1 state if residing in lieu of the excitons in STE2 state:

$$U_{exc} = \tfrac{1}{2} E_{gap}(\alpha_{exc}/\kappa)^2 \sum R_{ij}^{-6} \qquad (9)$$

Taking the ratio of (8) to (9) we get

$$U_{vib} / U_{exc} = [1 - (E_{gap}/4E_{JT})^2]^2 \exp(2E_{JT}/\hbar\omega) \qquad (10)$$

It is the large exponential term (reciprocal Holstein reduction factor) that makes $U_{vib} / U_{exc} \gg 1$. We also note that for small polaronic excitons $4E_{JT} > E_{gap}$.

## 3. Ball in crystalline matrix

Experimentally, luminous balls have been observed as clouds of excitons formed in Ge crystals under laser excitation.[10] The exciton concentration and binding energy have been estimated at $2\times10^{17}$ /cc and 2 meV, respectively. Unfortunately we cannot identify the exciton state attained in these experiments, whether STE1 or STE2. Under similar conditions the STE2 state might have produced an excessive binding energy, as shown in Table I of Ref. [5].

As far as the STE2 aggregation processes in an alkali halide matrix are concerned, they may prove essential before the subsequent exciton conversion to the DP state which splits the electron and hole clouds to bring about the disintegration of the exciton. As disintegration occurs on a massive scale, the anion sub-lattice turns into a lattice of F centers coexisting with the remaining alkali sub-lattice. (The F center is a halogen vacancy having trapped an electron). At the same time, the halogen atoms find themselves dissolved in the alkali ion – F center frame similar to what is happening in ionic superconductors. Halogen interstitials may subsequently aggregate to form halogen molecules and molecular ions. On splitting the electron & hole clouds, the colossal polarizability is transformed into F center and interstitial halogen polarizabilities.

## 4. Atmospheric ball

In any event, the ball lightning seems to be a rare atmospheric occurrence. Nevertheless, we shall single out some general properties of the real ball in the light of a fancy ball formed by a specific though fairly common material.

### 4.1. The ammonia molecule story

We shall first consider in some detail the case of a molecule that may happen to play a crucial role in our understanding of the ball lightning. As stated above, ammonia is a textbook example of broken-inversion-symmetry molecular cluster.[8,11] Ammonia may be suggested to form photochemically in the atmosphere during huge linear discharges, possibly by way of $6H_2O + 2N_2 \rightarrow 4NH_3 + 3O_2$. Ammonia is the case of a $D_{3h}$ symmetry $NH_3$ cluster in the unstable planar configuration which is stabilized by the nitrogen moving out of the $H_3$ plane to form a triangular pyramid with N residing at its vertex. Moving N out of the $H_3$ plane breaks the inversion symmetry of the planar cluster. The asymmetric $NH_3$ cluster possesses 2 distinct equilibrium modifications, left-handed and right-handed, with corresponding inversion dipoles equal in magnitude but directed oppositely to each other. At ambient temperature the system makes flip-flop jumps from one configuration to the other one as nitrogen passes easily through the hydrogen plane reverting the molecule. Because of these jumps the resultant electric dipole smears. Nevertheless, a system of ammonia molecules is polarizable electrically which gives rise to the Van-der-Waals intermolecular coupling. As we have seen in Section 2, the vibronic mixing may build up a great deal of polarizability resulting in a colossal VdW coupling.

Literature calculations of adiabatic potentials of $NH_3$ molecules show that the instability of the inversion-symmetric planar configuration in $^1A_1$' ground-state is lifted through the mixing of $^1A_1'$ with the second higher-lying $^1A_2''$ excited state by an $A_2''$ mode coordinate of the $D_{3h}$ point group.[8] From the data tabulated therein we get $E_{JT}$ = 0.16 eV at $E_{gap}$ = 5 eV (first excited state) and $E_{JT}$ = 5.16 eV at $E_{gap}$ = 14 eV (second excited state) relative to mixing with the ground state. We see that the small-polaron criterion $4E_{JT}/E_{gap} \gg 1$ is only met for the ground-state to second-excited state mixing which lifts the planar instability in the ground state. The stiffness in any state is K = 0.427×10$^5$ dyn/cm = 2.67 eV/Å$^2$. In so far as the $A_1'$ and $A_2''$ modes are in-phase hydrogen vibrations along with an counter-phase nitrogen vibration,[8] the common oscillator mass is M = $[3/M_H + 1/M_N]^{-1}$ ~ 1/3 and we arrive at a bare mode frequency of ω = 2.78×10$^{14}$ s$^{-1}$ = 0.18 eV. Using these data we calculate a Holstein reduction factor of exp(-2$E_{JT}/\hbar\omega$) = 1.26×10$^{-25}$ and ΔE = 1.76×10$^{-24}$ eV. Setting $p_{12}$ = 1 eÅ we get a huge vibronic polarizability of α = 6.52×10$^{-12}$ cm$^3$ = 6.52×10$^{12}$ Å$^3$. Using that estimate we calculate $U_{VdW}$ = 76 meV at $R_{ij}$ = 2 Å for κ = 1.5. This is more than 30 times higher than the exciton binding energy in Ge! The ammonia adiabatic potentials along the $A_2''$ mode coordinate are shown in Figure 2, as calculated from the literature data.[8] These potentials are computed using the equations:

$$E_{AD}(Q) = \tfrac{1}{2}\{KQ^2 \pm \sqrt{(4G^2Q^2 + E_{gap}^2)}\} \tag{11}$$

with parameters as follows: G = 5.24 eV/Å (electron-mode coupling constant), K = 2.67 eV/Å$^2$ (stiffness), $E_{gap}$ = 14 eV (gap energy). Q is the coupled mode coordinate.

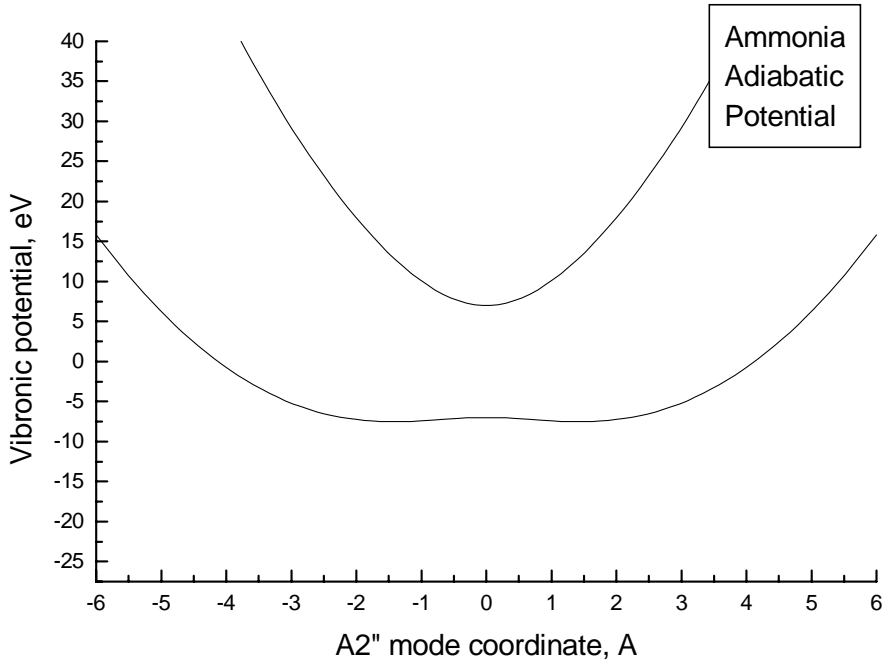

Figure 2.

Ammonia adiabatic potentials along the $A_2''$ mode coordinate ($D_{3h}$ point group) rendering unstable the planar configuration of the $NH_3$ molecule. Atomic displacements in the umbrella mode: H↓-H↓-H↓-N↑. The instability is lifted as the N atom is pushed firmly out of the $H_3$ plane to reside at the vertex of a regular $NH_3$ pyramid. The two pyramidal configurations, left-handed and right-handed, pertain to the two lateral minima of the ground-state potential. The position of a lower-lying excited state in which the N atom is in a Rydberg state is intermediate.

Similar adiabatic potentials have been obtained using eq.(10) for $NH_3^+$ (G = 3.27 eV/Å, K = 2,67 eV/Å$^2$, $E_{gap}$ = 12 eV) and $CH_4$ (G = 4.36 eV/Å, K = 2,67 eV/Å$^2$, $E_{gap}$ = 11 eV). [8] In all these cases the dipolar double-well instability has proved inherent to the ground electronic state, while the excited electronic state has always been single-well.

However, for a vibronic exciton phase to form at all, the stringent requirement is that the surrogate molecular cluster should possess a double-well dependence on a mixing mode coordinate in an *excited electronic state*. Unfortunately, this is not the case of the ammonia molecule which has been seen to have a double-well *ground electronic state* along the $A_1'$ mode coordinate, as well as two single-well $A_2''$ excited states.[8,11,12] In other words, the higher symmetry planar $NH_3$ configuration is unstable in ground state and stable in the excited states. Consequently, ammonia may build up a vibronic phase only in ground state, while its excitonic states will rather be STE1. Accordingly, ammonia may only build up dark cohesive clouds, while the ball-lightning clouds are expected to be luminous. Alternatively, it is conceivable considering the ammonia cluster as a precursor to the luminous ball. The *indirect* ball formation scenario is that while a #1 linear lightning builds up a supersaturated ammonia gas phase which condenses to a dark ball, a #2 lightning excites that ball to turn it luminous. This may be the case, provided the two discharges are well spaced in time to let the dark ball condense. Once turned luminous by the second discharge, the ball will ultimately disintegrate into a gas of ammonia molecules in a matter of a few seconds.

### 4.2. The surrogate molecule

For the *direct* formation of a luminous ball, we shall consider a surrogate molecular cluster of presently unspecified nature which can be photoexcited to build up a vibronic exciton phase and subsequently condense to an exciton cloud better known as *ball lightning*. This implies that two excited states of the surrogate cluster mix vibronically to lift an instability inherent to one of them by producing a double-well adiabatic potential therein. Examples of excited state instabilities are provided by defect formation processes via self-trapped excitons, 2p-2s excited states mixing at F centers in alkali halides, etc

It is tempting to propose the following sequence of fundamental processes leding to a ball: A huge linear-lightning discharge produces an amount of surrogate molecules in the excited electronic state. The main part of the excited molecules produce vibronic excitons which bind by way of the collosal VdW coupling leading to an excited surrogate gas phase. If the gas state is supersaturated, there will be excited surrogate condensation into clouds, possibly one cloud. Condensation will cease once the supersaturation has been depleted resulting in one single stable ball. Ultimately the surrogate vibronic excitons will relax radiatively to the ground electronic state emitting light in the orange region of the visible spectrum giving the luminous appearance of the ball formed. The observations suggest that there are both fluorescence and phosphorescence coming out of the ball which is understandable if a more complex tunneling transfer is involved. The condensed ball cloud will undoubtedly spill once the luminous emission has ceased, as the molecules will have relaxed to their ground states and the colossal binding will have vanished.

The nature of the surrogate molecules is not known, at least for the time being. Besides $NH_3$, two likely species to be produced during a linear discharge are $NH_3^+$ and $CH_4$. As told above, adiabatic energy calculations carried out for both of them show a dynamically unstable ground state and a stable excited state in the centrosymmetric configuration.[8] Both instabilities are lifted by the central atom moving out of the centrosymmetric point. Therefore these species are not fit for producing vibronic excitons, as well as $NH_3$. It will be interesting to list other reagents and photochemical products of a linear discharge for studying their adiabatic potentials. We can guess of $H_2O$, $CO_2$, $O_3$ and even of $SiO_2$ in so far as linear discharges often hit the silica rich ground.

The low shear displacement modulus leading to the high flexibility of the ball material has been discussed by Gilman.[1] We assume that his conclusions apply to any VdW binding whatever its specific nature.

## 5. Conclusion

In conclusion, we believe Gilman's suggestion for a highly cohesive ball state based on colossal Van der Waals attraction is basically correct. Herein we explored the alternative of a colossal Van der Waals coupling arising from Holstein's gap narrowing in small polaronic exciton systems. We considered a few illuminating cases and proposed a general scenario for the processes leading to ball formation. However, the basic surrogate substance for a ball matter remains unidentified. Nevertheless, we guess the present study may set the frames of future investigations into the problem. It will be essential to find if the ball formation proceeds directly to a luminous ball or indirectly via a dark precursor.

In this respect, we consider the vibronic coupling at the N atom excited Rydberg state, whose position is intermediate in Figure 2. Calculations have shown this state not to mix with the ground state of the ammonia molecule.[8] Instead, the next higher lying excited state has been found to mix strongly with the ground state. We agree but the Rydberg state not contributing to the chemical bond at the $NH_3$ molecule, it should, nevertheless, mix with the next higher-lying bonding state along the umbrella mode coordinate, for, otherwise, the system should disintegrate once in a Rydberg state. Strong mixing with the bonding state will produce a double-well character for the adiabatic potential of the Rydberg state. Inasmuch as this state is only 9 eV below the next excited bonding state, it will take a coupling constant $G \geq 0.87$ eV/Å to produce the strong mixing. We therefore appeal for improved numerical calculations of the $NH_3$ potentials.